\documentclass{article}%
\usepackage{placeins}
\usepackage{pbox}
\usepackage{graphicx}
\usepackage{amsmath}
\usepackage{amssymb}
\usepackage[margin=1in]{geometry}
\usepackage{verbatim}
\usepackage{float}
\usepackage{amsfonts}%
\providecommand{\U}[1]{\protect\rule{.1in}{.1in}}
\newtheorem{theorem}{Theorem}[section]

\newtheorem{lem}[theorem]{Lemma}
\newtheorem{rem}[theorem]{Remark}
\newtheorem{thm}[theorem]{Theorem}
\newtheorem{conj}[theorem]{Conjecture}
\newtheorem{openq}[theorem]{Open question}
\newtheorem{prop}[theorem]{Proposition}

\newenvironment{enumerate2}{
\begin{enumerate}
\setlength{\itemsep}{1pt}
\setlength{\parskip}{0pt}
\setlength{\parsep}{0pt}
}{\end{enumerate}}
\begin{document}

\title{Maximizing algebraic connectivity for certain families of graphs}
\author{ \renewcommand{\thefootnote}{\alph{footnote}} 
T.~Kolokolnikov \\
\\
{\small \emph{Department of Mathematics and Statistics, Dalhousie
University, Halifax, Nova Scotia, B3H 3J5 Canada}}\\
[3ex]}
\date{}
\maketitle

\begin{abstract}
We investigate the bounds on algebraic connectivity of graphs subject to
constraints on the number of edges, vertices, and topology. We show that the
algebraic connectivity for any tree on $n$ vertices and with maximum degree
$d$ is bounded above by $2\left(  d-2\right)  \frac{1}{n}+O\left(  \frac{\ln
n}{n^{2}}\right)  .$ We then investigate upper bounds on algebraic
connectivity for cubic graphs. We show that algebraic connectivity of a cubic
graph of girth $g$ is bounded above by $3-2^{3/2}\cos\left(  \pi/\left\lfloor
g/2\right\rfloor \right)  ,$ which is an improvement over the bound found by
Nilli [A. Nilli, Electron. J. Combin., 11(9), 2004]. Finally, we propose
several conjectures and open questions.
\end{abstract}



\section{Introduction}

This paper is motivated by the following question:\ among all possible
networks connecting $n$ nodes, and subject to a specified resource or topology
constraints, which one is the most effective at diffusing the flow of
information?\ We are interested in the case where the network is undirected
and all non-zero edges have the same weight.

One of the simplest ways of modelling the information flow in a network is the
linear consensus model, which is widely used in control theory
\cite{olfati2004consensus}:%
\begin{equation}
\frac{du_{j}}{dt}=\sum_{j\neq i}e_{ij}\left(  u_{i}-u_{j}\right)  .
\label{ode}%
\end{equation}
Here $e_{ij}$ denote edge weights between nodes $i,j$ and $u_{j}$ is the
\textquotedblleft load\textquotedblright\ at node $j$; the information flows
from $i$ to $j$ in proportion to the load differential between the nodes;
$e_{ij}=1$ if $i$ and $j$ are joined by an edge and is zero otherwise. For
large $t$ the solution to (\ref{ode}) is given by $u(t)\sim\bar{u}%
+Ce^{-\lambda_{2}t},$ where $\bar{u}$ is consensus (average) state and
$\lambda_{2}$ is the second smallest eigenvalue of the graph Laplacian matrix
$L=D-A$ where $A$ is the adjacency matrix and $D$ is the degree matrix (the
smallest eigenvalue of $L$ is zero and $\lambda_{2}>0$ if and only if the
network is connected). The eigenvalue $\lambda_{2}$ is often called the
\emph{algebraic connectivity }of the graph \cite{fiedler1973algebraic}, and
roughly, the larger $\lambda_{2},$ the faster $u$ diffuses to its consensus
state. In this sense, the \textquotedblleft optimal\textquotedblright\ network
is the one which maximizes the algebraic connectivity, subject to given
constraints. This leads to the following question.

\textbf{Question:\ }\emph{Which graphs maximize the algebraic connectivity,
given a set of constraints on the number of vertices, edges, maximum degree,
and graph topology?}

This and related questions arise in many\ diverse areas, including optimal
network topologies \cite{donetti2006optimal}; scheduling and network coding
\cite{koetter2003algebraic}; experimental design \cite{boyd2004fastest,
chung2012experimental}, diffusion in small world networks
\cite{olfati2005ultrafast, delre2007diffusion}, synchronization in complex
networks \cite{arenas2008synchronization}, and ranking algorithms
\cite{osting2013enhanced, osting2012optimal}. There is also a close link to
expander graphs and Ramanujan graphs. These are graphs with \textquotedblleft
high\textquotedblright\ algebraic connectivity in some sense. See recent
reviews \cite{hoory2006expander, lubotzky2012expander} and references therein.
A nice recent survey on algebraic connectivity is \cite{de2007old}.

In general, the problem of finding the optimal graph given $m$ edges and $n$
vertices is known to be NP-complete \cite{mosk2008maximum}. Despite this fact,
several simple heuristics exist that can be used to obtain a graph with
reasonably large algebraic connectivity \cite{ghosh2006growing,
wang2008algebraic}. See also \cite{belhaiza2005variable} for some results for
almost-complete graphs, where $m$ is close to $n\left(  n-1\right)  /2.$ In
\cite{fallat1998extremizing, wang2010graphs}, the question of optimizing
algebraic connectivity with respect to graph diameter was studied.

\begin{figure}[tb]
\begin{center}
\includegraphics[width=1\textwidth]{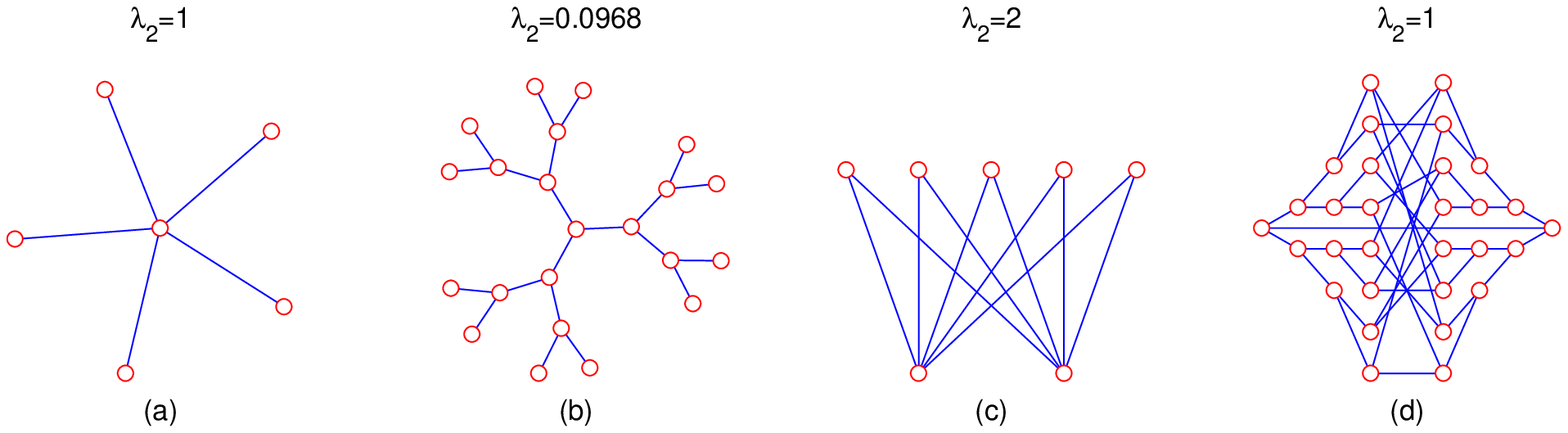}
\end{center}
\caption{(a) The star, which is the
maximizer of algebraic connectivity for all trees. (b) Maximally-balanced tree of
degree at most $d$ with $n=\frac{d(d-1)^{K}-2}{d-2}$ vertices (here,
$n=22,d=3,K=3$). (c) The complete bipartite graph $K_{2,n-2}$, which is a
conjectured maximizer for all graphs of $n$ vertices with $m=2(n-2)$ edges.
(d)\ The Tutte 8-cage, which is the conjectured maximizer for the cubic graphs
with 30 vertices.}%
\label{fig:canonical}%
\end{figure}

In this paper we are concerned with the regime where the number of edges $m$
grows in proportion to the number of vertices $n,$ so that the graph is
relatively sparse. In particular, a random Erdos-Renyei graph with $O(n)$
edges is well known to be disconnected with high probability as $n\rightarrow
\infty,$ so for such a graph, $\lambda_{2}=0$ almost surely \cite{erdos,
alon2004probabilistic}, and as such, random graphs are not good optimizers in
this regime. The smallest value for $m$ for which the graph is connected is
$m=n-1,$ in which case any connected graph is a tree (for disconnected graphs,
$\lambda_{2}=0$ so we only consider connected case). Without a degree
restriction, the star, which is a tree having a single root and $n-1$ leafs (see
Figure \ref{fig:canonical}(a)), is the unique optimizer of algebraic
connectivity among all trees of $n$ vertices, with $\lambda_{2}=1$ when
$n\geq3$ \cite{grone1990laplacian, de2007old}. However, many trees of
importance to applications have a degree restriction. For example, decision or
binary trees have degree at most 3. Another important example are trees
representing neuronal dendrites \cite{saito2011phase}, which consist of mostly
degree two vertices with an occasional degree 3 vertex (see
\cite{saito2011phase} for further details). This motivates the following
question.{}

\begin{openq}
\label{openq2}Among all trees with $n$ vertices with maximal vertex degree
$d,$ which tree maximizes the algebraic connectivity?
\end{openq}

In this paper we give the following partial answer to this question:

\begin{thm}
\label{thm:tree}Let $T$ be any tree with $n$ vertices and maximum degree $d.$
Then $\lambda_{2}\left(  T\right)  \leq2\left(  d-2\right)  \frac{1}%
{n}+O\left(  \frac{\ln n}{n^{2}}\right)  $ as $n\rightarrow\infty$ for fixed
$d.$
\end{thm}

A bound without the $O$ notation (valid even when $n=O(1)$), is given in
(\ref{precise1}).

A well-known \textquotedblleft basic\textquotedblright\ upper bound for
algebraic connectivity for any tree is $\lambda_{2}(T)\leq2-2\cos(\frac{\pi
}{D+1}),$ where $D$ is the diameter of $T$ \cite{de2007old}, and can be
obtained by \textquotedblleft pruning\textquotedblright\ any branches that are
not along the longest path of the tree. This bound is attained for both the
star graph and the path graph. However, in general, it is far from optimal
when there is a restriction on the maximal degree of a tree. Among trees of
maximal degree $d,$ one has $n\leq\frac{d\left(  d-1\right)  ^{D/2}-2}{d-2}$
(the equality is achieved only for a maximally balanced tree such as shown in
Figure \ref{fig:canonical}(b). A maximally balanced tree is a tree whose leafs
are all at the same distance from a root vertex and whose non-leaf vertices
all have the same degree). For fixed $d$ and large $n,$ this yields $D\geq
O(\ln n)$ so that the \textquotedblleft basic\textquotedblright\ bound is
$\lambda_{2}(T)\leq O(1/\ln^{2}(n)),$ which is much worse than the $O(1/n)$
bound of Theorem \ref{thm:tree}. The lower bound for the algebraic
connectivity of any tree of $n$ vertices is attained by the path graph for
which $\lambda_{2}=2-2\cos(\frac{\pi}{n})=O(\frac{1}{n^{2}})$, so that in
general, $O(1/n^{2})\leq\lambda_{2}\leq O(1/n)$.

The algebraic connectivity of a maximally balanced tree such as shown in
Figure \ref{fig:canonical}(b) can be determined explicitly, as was done for
example in \cite{molitierno2000tight, rojo2002spectrum, rojo2006tight}. It was
found that $\lambda_{2}\sim\frac{d}{d-1}\left(  d-2\right)  \frac{1}{n}$ as
$n\rightarrow\infty$ for such a tree. So the bound in Theorem \ref{thm:tree}
is not optimal; in fact we conjecture that $\frac{d}{d-1}\left(  d-2\right)
\frac{1}{n}$ is the asymptotically optimal upper bound as $n\to \infty$. 
See Section \ref{sec:discuss} for
further discussion and a related conjecture.

In Section \ref{sec:cubic} we explore optimal cubic (i.e. 3-regular)\ graphs,
which have $m=3n/2$ edges. We are motivated by the following question.

\begin{openq}
Among all cubic (i.e. 3-regular)\ graphs with $n$ vertices, which one
maximizes the algebraic connectivity?
\end{openq}

Regular graphs appear in numerous applications where having high connectivity
is important. It is well known that the expected algebraic connectivity of a
random cubic graph is $\lambda_{2}\sim3-2\sqrt{2}+O(1/\ln(n))$ as
$n\rightarrow\infty$ (see \cite{mckay1981expected, alon1986eigenvalues,
broder1987second, friedman1991second}). So unlike the case of trees of maximum
degree 3, the maximum possible connectivity of a cubic graph is bounded away
from zero. One of the applications of this fact is that a random cubic graph
is an expander graph with very high probability \cite{wormald1999models,
murty2003ramanujan}.

The best known bound for $\lambda_{2}$ was obtained by Nilli in
\cite{nilli2004tight}. He showed that for any cubic graph, $\lambda_{2}%
\leq3-2\sqrt{2}\cos\left(  2\pi/D\right)  $ where $D$ is its diameter. However
so far, there is no example of a cubic graph that we know of, which actually
attains this bound. In Section \ref{sec:discuss} we suggest a possible optimal
bound when $n=2^{K}-2,$ which is tighter than Nilli's bound, and which is
achieved at least for $n=6, 14,30$ and $126$. This is discussed in Conjecture
\ref{conj:cubic}. Related to this conjecture, we prove the following result.

\begin{thm}
\label{thm:cubic}Suppose that a cubic graph $G$ of $n$ edges has girth $g$.
Then $\lambda_{2}\left(  G\right)  \leq3-2\sqrt{2}\cos\left(  \pi/\left\lfloor
g/2\right\rfloor \right)  .$
\end{thm}

For some graphs, this bound is actually achieved; see Figure
\ref{fig:canonical}(d)\ and Section \ref{sec:discuss}. As shown in Remark
\ref{remark:cubic} below, the bound of Theorem \ref{thm:cubic} is better the
result obtained by Nilli in \cite{nilli2004tight}, which is $\lambda
_{2}\left(  G\right)  \leq3-2\sqrt{2}\cos\left(  2\pi/\left\lfloor
g/2\right\rfloor \right)  .$

Finally, in Section \ref{sec:discuss}, we discuss some numerical results, open
questions and several conjectures, including the following conjecture:

\begin{conj}
\label{conj:K2}Among all graphs with exactly $n$ vertices and $m=2\left(
n-2\right)  $ edges, a graph which maximizes the algebraic connectivity is the
complete bipartite graph $K_{2,n-2}$ (see Figure \ref{fig:canonical}(c)), with
$\lambda_{2}(K_{2,n-2})=2.$
\end{conj}

\section{Trees\label{sec:trees}}

In this Section we prove Theorem \ref{thm:tree}. We recall the alternative
definition of $\lambda_{2}$ for a graph $G$ on $n$ vertices using the Rayleigh
quotient \cite{de2007old},
\begin{equation}
\lambda_{2}=\min_{\substack{x\in%
\mathbb{R}
^{n}\text{ subject to}\\x_{1}+\cdots+x_{n}=0}}\frac{\sum_{(i,j)\in
E(G)}\left(  x_{i}-x_{j}\right)  ^{2}}{\sum_{j=1}^{n}x_{j}^{2}}.
\label{lambda2}%
\end{equation}
We first need the following concept of a \textquotedblleft
modified\textquotedblright\ Laplacian eigenvalue. Given a graph $G$ and a
vertex $r\in V(G),$ define%
\begin{equation}
\tilde{\lambda}(G,r):=\min_{x\in%
\mathbb{R}
^{n}}\frac{x_{r}^{2}+\sum_{\left(  i,j\right)  \in E(G)}\left(  x_{j}%
-x_{i}\right)  ^{2}}{\sum_{j=1}^{n}x_{j}^{2}} \label{lambdat}%
\end{equation}
An alternative definition is that $\tilde{\lambda}$ is the smallest eigenvalue
of the eigenvalue problem%
\begin{equation}
\left\{
\begin{array}
[c]{c}%
\tilde{\lambda}x_{j}=\sum_{\left(  i,j\right)  \in E(G)}\left(  x_{j}%
-x_{i}\right)  ,\ \ \text{if }j\neq r\\
\tilde{\lambda}x_{j}=x_{j}+\sum_{\left(  i,j\right)  \in E(G)}\left(
x_{j}-x_{i}\right)  ,\ \text{if}\ j=r
\end{array}
\right.
\end{equation}
The proof of Theorem \ref{thm:tree} relies on the following three lemmas:

\begin{lem}
\label{Lemma:lamtilde}Let $T$ be a tree with $n$ vertices each of
degree at most $d,$ and whose root $r$ has degree at most $d-1$. Then
$\tilde{\lambda}(T,r)\leq\frac{d-2}{d-1}\frac{1}{n}+O\left(  \frac{\ln
n}{n^{2}}\right)  .$
\end{lem}

\begin{lem}
\label{Lemma:max}Given a graph $G,$ and a vertex $v$ with at least two edges
$\left(  v,u\right)  $ and $\left(  v,w\right)$ such that removing $v$
separates $G$ into at least two or more disjoint subgraphs $G_{1},G_{2}%
\ldots,$ such that $u\in V(G_{1})$ and $w\in V(G_{2}).$ Then%
\[
\lambda_{2}(T)\leq\max\left(  \tilde{\lambda}\left(  G_{1},u\right)
,\tilde{\lambda}\left(  G_{2},w\right)  \right)  .
\]

\end{lem}

\begin{lem}
\label{Lemma:middle}Let $T$ be a tree with $n$ vertices and of
maximal degree $d.$ Then there exists a vertex $v\in V(T)$ such that removing
$v$ and its associated edges separates $T$ into subtrees such that at least
two of these subtrees have at least $\frac{n-2}{2(d-1)}$ vertices.
\end{lem}

\textbf{Proof of Lemma \ref{Lemma:lamtilde}.} Choose unique positive integers
$m$ and $K$ such that
\begin{align}
n  &  =1+(d-1)+(d-1)^{2}+\cdots+(d-1)^{K-1}+m,\ \ \ \ \text{with }0\leq
m<(d-1)^{K}\nonumber\\
&  =\frac{(d-1)^{K}-1}{d-2}+m. \label{mK}%
\end{align}
Sort the vertices according to their distance from the root, from smallest to
largest. After sorting them, let $V_{1}$ be the set containing the first
vertex in the list, i.e. root vertex; let $V_{2}$ contain the next $d-1$
vertices; let $V_{3}$ contain the next $(d-1)^{2}$ vertices and so on up to
$V_{K}$ which contains $\left(  d-1\right)  ^{K-1}$ vertices$,$ and with
$V_{K+1}$ containing the remaining $m$ vertices. For vertex $j\in V_{k}$,
assign a weight $x_{j}=1-\left(  \frac{1}{d-1}\right)  ^{k}.$

For a non-root vertex $j\in V(T), j\ne r,$ let $parent(j)\in V(T)$ denote its
parent, that is the neighbouring vertex that is closer to the root $r$. We then
have%
\begin{align*}
x_{r}^{2}+\sum_{\left(  i,j\right)  \in E(T)}\left(  x_{j}-x_{i}\right)  ^{2}
&  =x_{r}^{2}+\sum_{j\in V(T),\ j\neq r}\left(  x_{j}-x_{parent(j)}\right)
^{2}\\
&  =x_{r}^{2}+\sum_{k=2}^{K+1}\sum_{j\in V_{k}}\left(  x_{j}-x_{parent(j)}%
\right)  ^{2}%
\end{align*}
Moreover, if $j\in V_{k}$ with $k>1,$ then either $parent(j)\in V_{k}$ or else
$parent(j)\in V_{k-1}.$ In both cases, we have%
\[
\left(  x_{j}-x_{parent(j)}\right)  ^{2}\leq\left(  \frac{1}{(d-1)^{k}}%
-\frac{1}{\left(  d-1\right)  ^{k-1}}\right)  ^{2}=\frac{\left(  d-2\right)
^{2}}{\left(  d-1\right)  ^{2k}}%
\]
so that%
\begin{align}
x_{r}^{2}+\sum_{\left(  i,j\right)  \in E(T)}\left(  x_{j}-x_{i}\right)  ^{2}
&  \leq\left(  \frac{d-2}{d-1}\right)  ^{2}+\sum_{k=2}^{K}(d-1)^{k-1}%
\frac{\left(  d-2\right)  ^{2}}{\left(  d-1\right)  ^{2k}}+m\frac{\left(
d-2\right)  ^{2}}{\left(  d-1\right)  ^{2\left(  k+1\right)  }}\label{nav1}\\
&  =\frac{d-2}{d-1}-\frac{d-2}{(d-1)^{K}}+m\frac{\left(  d-2\right)  ^{2}%
}{\left(  d-1\right)  ^{2\left(  k+1\right)  }}\nonumber\\
&  \sim\frac{d-2}{d-1}+O(1/n).\nonumber
\end{align}

Similarly, we write%
\[
\sum_{i\in V(T)}x_{i}^{2}=\sum_{k=2}^{K}\sum_{j\in V_{k}}x_{j}^{2}.
\]
Moreover, for $j\in V_{k},$ we have $x_{j}=1-\left(  \frac{1}{d}\right)  ^{k}$
so that%
\begin{align}
\sum_{i\in V(T)}x_{i}^{2}  &  =\sum_{k=2}^{K}(d-1)^{k-1}\left(  1-\left(
\frac{1}{d-1}\right)  ^{k}\right)  ^{2}+m\left(  1-\left(  \frac{1}%
{d-1}\right)  ^{K+1}\right)  ^{2}\label{nav2}\\
&  =\left(  \frac{\left(  d-1\right)  ^{K}}{\left(  d-2\right)  }+m\right)
\left[  1+O\left(  \frac{K}{\left(  d-1\right)  ^{K}}\right)  \right]
\nonumber\\
&  =n\left(  1+O(K/n)\right) \nonumber
\end{align}
Therefore 
\[
\frac{x_{r}^{2}+\sum_{\left(  i,j\right)  \in E(T)}\left(  x_{j}-x_{i}\right)
^{2}}{\sum_{j=1}^{N}x_{j}^{2}}=\frac{d-2}{d-1}\frac{1}{n}+O(K/n^{2}).
\]
Moreover, note from definition (\ref{mK})\ of $m$ and $K$ that $K=O(\ln n)$ so
that $O(K/n^{2})=O(\left(  \ln n\right)  /n^{2}).$ Recalling the definition
(\ref{lambdat})\ of $\tilde{\lambda}$ completes the proof of the lemma.
$\blacksquare$

\begin{rem}
\emph{The }$O$\emph{ notation can be avoided by computing all the terms in
(\ref{nav1})\ and (\ref{nav2}). Setting }$m=0,$\emph{ we then obtain the upper
bound without the }$O$\emph{ notation,}%
\begin{equation}
\tilde{\lambda}\leq\frac{\left(  d-2\right)  ^{2}}{(d-1)^{K+1}}\left(
\frac{1-\frac{1}{(d-1)^{K-1}}}{1-\frac{2(K-1)}{\left(  d-1\right)  ^{K}%
}(d-2)-\frac{d-1-\left(  d-1\right)  ^{-2}}{\left(  d-1\right)  ^{K}}-\frac
{1}{\left(  d-1\right)  ^{2K+1}}}\right)  .\label{precise2}%
\end{equation}
\emph{The same bound is valid even if }$m>0$\emph{, because appending leafs to a
tree only decreases }$\tilde{\lambda}$\emph{ (see \cite{de2007old}). The bound
(\ref{precise2})\ is very close (but not identical)\ to the upper bound as was
obtained for Bethe trees with }$k=K+1$\emph{ levels in
\cite{molitierno2000tight, rojo2006tight} using a related method.}
\end{rem}

\textbf{Proof of Lemma \ref{Lemma:max}. }Let $x$ be the eigenvector
corresponding to $\tilde{\lambda}(G_{1},u)$ and $y$ be the eigenvector
corresponding to $\tilde{\lambda}(G_{2},w),$ so that $x_{j}=0$ for all
$j\notin V(G_{1})$ and similarly $y_{j}=0$ for all $j\notin V(G_{2}).$

Consider any linear combination $z=\alpha x+\beta y.$ Note that $z_{v}=0$ and
we have%

\[
\sum_{(i,j)\in E(G)}\left(  z_{i}-z_{j}\right)  ^{2}=\alpha^{2}\left(
\sum_{(i,j)\in E(G_{1})}\left(  x_{i}-x_{j}\right)  ^{2}+x_{u}^{2}\right)
+\beta^{2}\left(  \sum_{(i,j)\in E(G_{2})}\left(  y_{i}-y_{j}\right)
^{2}+y_{w}^{2}\right)  .
\]
Moreover by orthogonality, we have $\left\vert z\right\vert ^{2}=\alpha
^{2}\left\vert x\right\vert ^{2}+\beta^{2}\left\vert y\right\vert ^{2}.$
Define%
\[
R_{1}(x):=\dfrac{\sum_{(i,j)\in E(G_{1})}\left(  x_{i}-x_{j}\right)
^{2}+x_{u}^{2}}{\sum_{i\in V(G_{1})}x_{i}^{2}},\ \ \ \ \ R_{1}(y):=\dfrac
{\sum_{(i,j)\in E(G_{2})}\left(  y_{i}-y_{j}\right)  ^{2}+y_{w}^{2}}%
{\sum_{i\in V(G_{2})}y_{i}^{2}}%
\]
so that $\tilde{\lambda}(G_{1},u)=R_{1}(x),\ $ $\tilde{\lambda}(G_{2}%
,w)=R_{2}(y).$ We have%
\begin{align*}
\frac{\sum_{(i,j)\in E(G)}\left(  z_{i}-z_{j}\right)  ^{2}}{\left\vert
z\right\vert ^{2}}  &  =\dfrac{\alpha\left(  \sum_{(i,j)\in E(G_{1})}\left(
x_{i}-x_{j}\right)  ^{2}+x_{u}^{2}\right)  +\beta\left(  \sum_{(i,j)\in
E(G_{2})}\left(  y_{i}-y_{j}\right)  ^{2}+y_{w}^{2}\right)  }{\alpha
^{2}\left\vert x\right\vert ^{2}+\beta^{2}\left\vert y\right\vert ^{2}}\\
&  =R_{1}(x)\frac{\alpha^{2}\left\vert x\right\vert ^{2}}{\alpha^{2}\left\vert
x\right\vert ^{2}+\beta^{2}\left\vert y\right\vert ^{2}}+R_{2}(y)\frac
{\beta^{2}\left\vert y\right\vert ^{2}}{\alpha^{2}\left\vert x\right\vert
^{2}+\beta^{2}\left\vert y\right\vert ^{2}}\\
&  \leq\max(R_{1}(x),R_{2}(y))\\
&  \leq\max\left(  \tilde{\lambda}(G_{1},u),\tilde{\lambda}(G_{2},w)\right)  .
\end{align*}
Now choose $\alpha,\beta$ such that $\sum_{i\in V(G)}z_{i}=0$. That is, take
$\frac{\alpha}{\beta}=-\left(  \sum_{i\in V(G_{2})}y_{i}\right)  /\left(
\sum_{i\in V(G_{1})}x_{i}\right)  $ as long as $\sum_{i\in V(G_{1})}x_{i}%
\neq0;$ in the contrary case take $\alpha=1,\beta=0.$ Then from the
definition (\ref{lambda2})\ of $\lambda_{2},$ we get
\[
\lambda_{2}\leq\frac{\sum_{(i,j)\in E(G)}\left(  z_{i}-z_{j}\right)  ^{2}%
}{\left\vert z\right\vert ^{2}}\leq\max\left(  \tilde{\lambda}(G_{1}%
,u),\tilde{\lambda}(G_{2},w)\right)
\]
which concludes the proof. $\blacksquare$

We note that an alternative proof of Lemma \ref{Lemma:max} can be given using
the mini-max definition of $\lambda_{2}$, as done by Nilli in
\cite{nilli2004tight}.

\textbf{Proof of Lemma \ref{Lemma:middle}}. The algorithm to find $v$ is
simple:\ start with an arbitrary vertex $v_{0}$ in $T.$ Choose a neighbour
$v_{1}$ of $v_{0}$ which belongs to the subtree with the largest number of
vertices, among all the subtrees that are obtained by deleting $v_{0}$ from
$T$ (in case of a tie, choose a vertex deterministically, e.g. the one with
the smallest index). Continue this process, obtaining a sequence of vertices
$v_{0},v_{1},v_{2},\ldots$ This sequence eventually settles to a two-cycle
$v,w,v,w,\ldots$ When this happens, consider the two subtrees obtained by
deleting the edge $(v,w),$ call them $T_{1}$ and $T_{2}.$ One of these tree,
say tree $T_{1}$ containing $v,$ has at least $n/2$ vertices. Upon deleting
$v$ from this tree, we get at most $d-1$ subtrees of $T_{1}$. So one of these
subtrees must have the size at least ($n/2-1)/(d-1)=\frac{n-2}{2(d-1)}$
vertices. But then the second tree $T_{2}$ containing $w$ must have at least
$\frac{n-2}{2(d-1)}$ vertices also, since it is the subtree that contains the
most vertices among all subtrees obtained by deleting $v.$ So $v$ is the
desired vertex. $\blacksquare$

We are now in position to prove the main theorem of this paper.

\textbf{Proof of Theorem \ref{thm:tree}}. Choose a vertex $v$ using Lemma
\ref{Lemma:middle}, which separates the tree into at least two subtrees whose
sizes are $n_{1},n_{2}\geq\frac{n-2}{2(d-1)}.$ Applying Lemmas \ref{Lemma:max}
and \ref{Lemma:lamtilde} to these subtrees we obtain%
\[
\lambda_{2}(T)\leq\frac{d-2}{d-1}\max\left(  \frac{1}{n_{1}},\frac{1}{n_{2}%
}\right)  +O\left(  \frac{\ln n}{n^{2}}\right)  \leq\frac{d-2}{d}\frac{2}%
{n}+O\left(  \frac{\ln n}{n^{2}}\right)  .\ \ \ \ \ \ \ \ \ \ \ \blacksquare
\]

\bigskip

The bound in Theorem \ref{thm:tree} can be written without the $O$ notation,
by replacing the estimate for $\tilde{\lambda}$ in Lemma \ref{Lemma:lamtilde}%
\ with the estimate (\ref{precise2}). To do this, choose $K$ in
(\ref{precise2})\ in such a way that the number of vertices in two subtrees
produced by Lemma \ref{Lemma:middle} is more than $\frac{\left(  d-1\right)
^{K}-1}{d-2}$ (formula (\ref{mK}) with $m=0).$ That is, choose $K$ such that
$\frac{\left(  d-1\right)  ^{K}-1}{d-2}\leq\frac{n-2}{2(d-1)}.$ We then obtain
an upper bound without the $O$ notation, namely that%

\begin{equation}
\lambda_{2}\leq\text{right hand side of (\ref{precise2}) with }K=\left\lfloor
\log_{d-1}\left(  1+\frac{\left(  d-2\right)  \left(  n-2\right)  }{2\left(
d-1\right)  }\right)  \right\rfloor . \label{precise1}%
\end{equation}

\section{Cubic graphs\label{sec:cubic}}

In this Section we give the proof of Theorem \ref{thm:cubic}. It is a direct
consequence of the following lemma.

\begin{lem}
\label{Lemma:Tk}Let $T_{K}$ be a graph consisting of two perfect binary trees
of height $K$ joined by an edge connecting their roots as illustrated here
(with $K=3$):
\begin{gather*}
\includegraphics[width=0.5\textwidth]{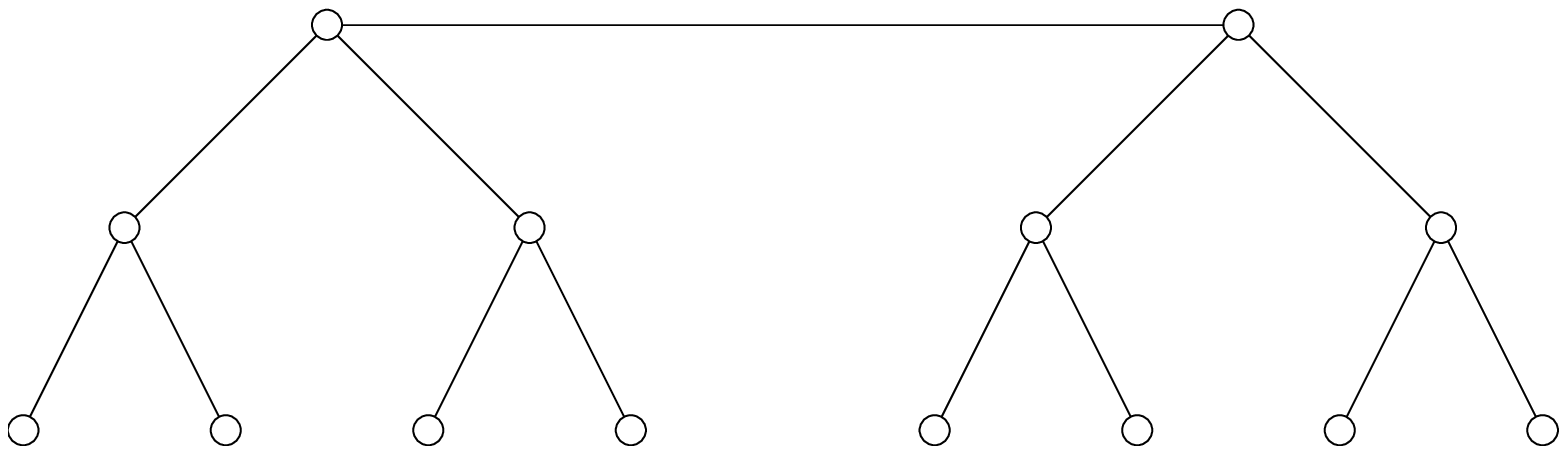}\\
K=3
\end{gather*}
Suppose that a cubic graph $G$ has $T_{K}$ as its subgraph. Then $\lambda
_{2}(G)\leq3-2^{3/2}\cos(\pi/K)$.
\end{lem}

Above, we defined the height $K$ of a perfect binary tree as one less than 
the distance from
any leaf to its root.

\begin{rem}
\label{remark:cubic} \emph{It was shown by Nilli \cite{nilli2004tight} that
for any cubic graph }$G,$\emph{ }$\lambda_{2}(G)\leq3-2^{3/2}\cos\left(
2\pi/D\right)  $\emph{ where }$D$\emph{ is the diameter of the graph. If }%
$G$\emph{ has }$T_{K}$\emph{ as its subgraph, then it has two vertices that
are separated by distance at least }$K$:\emph{ take the first vertex to be the
root of one of the two binary trees that make up }$T_{K}$\emph{ and take the
second vertex to be one of the leafs of the other subtree. So Nilli's bound
for the algebraic connectivity of }$G$\emph{ is }$3-2^{3/2}\cos(2\pi
/K)$\emph{. Thus, Lemma \ref{Lemma:Tk} is an improvement over Nilli's bound
for the case where }$T_{K}$\emph{ is a subgraph of }$G.$\emph{ Similarly, a
graph of girth }$g$\emph{ has diameter at least }$\left\lfloor
g/2\right\rfloor ,$\emph{ so that Nilli's bound is }$3-2^{3/2}\cos
(2\pi/\left\lfloor g/2\right\rfloor )$\emph{ which is worse than the result of
Theorem \ref{thm:cubic}.}
\end{rem}

\textbf{Proof.} Consider the following choice of weights $x_{j},j\in V(G)$:
for nodes at level $k$ on the right tree, assign weight $x_{j}=v_{k},$ where
$v_{k}$ will be specified below; for nodes at level $k$ on the left tree,
assign weight $x_{j}=-v_{k}$. For all other nodes, assign weight zero. With
this choice, the sum of all the weights is zero, so that $\left(  x_{1}%
,\ldots,x_{n}\right)  \perp\left(  1,1,\ldots,1\right)  $. Now consider any
leaf vertex of $T_{K}.$ It has three edges:\ one that connects it to its
parent, and two other edges that connect it to either another leaf whose
weight is $\pm v_{K}$ or to a vertex outside $T_{K}$ whose weight is zero.
Therefore if $\left(  v,w\right)  $ is an edge that connects $v$ to the
non-parent vertex $w$ and $a,b$ are the weights of $v$ and $w$ respectively,
then $\left(  a-b\right)  ^{2}\leq\left(  v_{K}-(-v_{K})\right)  ^{2}.$

It follows that $\lambda_{2}(G)$ bounded by any eigenvalue $\mu$ of the
eigenvalue problem
\begin{align}
\mu v_{1}  &  =\left(  v_{1}-(-v_{1})\right)  +2\left(  v_{1}-v_{2}\right)
;\label{tri1}\\
\mu v_{j}  &  =v_{j}-v_{j-1}+2\left(  v_{j}-v_{j+1}\right)  ,\ \ \ j=2\ldots
K-1\label{tri2}\\
\mu v_{K}  &  =v_{K}-v_{K-1}+2(v_{K}-(-v_{K})), \label{tri3}%
\end{align}
corresponding to the $K$ by $K$ matrix%
\begin{equation}
M=\left[
\begin{array}
[c]{cccccc}%
4 & -2 &  &  &  & \\
-1 & 3 & -2 &  &  & \\
& -1 & \ddots & \ddots &  & \\
&  & \ddots & 3 & -2 & \\
&  &  & -1 & 3 & -2\\
&  &  &  & -1 & 5
\end{array}
\right] \nonumber
\end{equation}
Similar types of Toeplitz matrices are well-known and occur in many related
problems, for example when computing eigenvalues of Bethe trees
\cite{molitierno2000tight, rojo2006tight}. For reader's convenience, here we
show directly that its eigenvalues are given by $\mu=3-2^{3/2}\cos(\pi k/K),$
$k=1\ldots K.$

We have the following self-consistent anzatz for the eigenvector:%
\begin{equation}
v_{j}=Az^{j}+B\left(  \frac{1}{2z}\right)  ^{j},
\end{equation}
where $A,B$ and $z$ are to be found. Then it is easy to check that $\mu
v_{j}=v_{j}-v_{j-1}+2\left(  v_{j}-v_{j+1}\right)  $ holds for any $j,A,B$
whenever%
\begin{equation}
\mu=3-2z-\frac{1}{z}.
\end{equation}
\ Write (\ref{tri1}) as
\[
\mu v_{1}=3v_{1}-2v_{2}-v_{0}+v_{0}+v_{1}.
\]
It follows that $v_{0}+v_{1}=0$ so that $A=-B.$ Similarly, from the last row
we obtain $2v_{k}+2v_{K+1}=0,$ which yields an equation for $z.$ After some
algebra, this equation simplifies to
\[
z^{-2K}=2^{K},\ \ z\neq\left(  1/2\right)  ^{1/2}.
\]
so that $z=\left(  \frac{1}{2}\right)  ^{1/2}e^{\frac{2\pi ij}{2K}}.$ The
choice $j=0$ corresponds to $v_{j}=0$ for all $j$ so this is not allowed. The
remaining choices are%
\[
\mu=3-2\sqrt{2}\cos(\pi j/K),\ \ \ j=1\ldots K
\]
The smallest eigenvalue among these corresponds to the choice $j=1,$ which is
precisely the bound of the lemma. $\blacksquare$

\textbf{Proof of Theorem \ref{thm:cubic}.} A cubic graph of girth $g$ has a
subtree $T_{K}$ as defined in Lemma \ref{Lemma:Tk}, where $K$ is any integer
at most $g/2.$ Applying Lemma \ref{Lemma:Tk} completes the proof.
$\blacksquare$

\section{Computer experiments, open questions, discussion\label{sec:discuss}}

\begin{figure}[ptb]
\begin{center}
\includegraphics[width=1\textwidth]{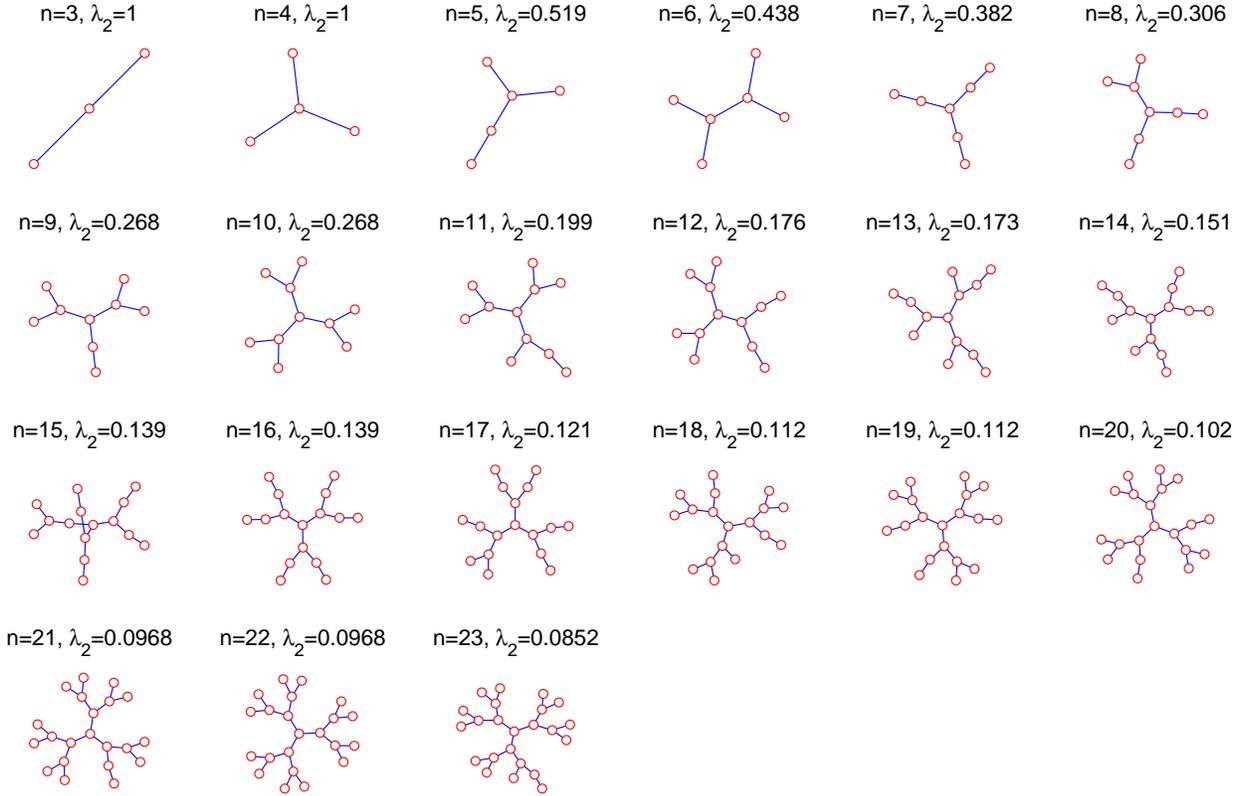}
\end{center}
\caption{Trees of degree 3 on $n$ vertices that have maximum possible the
algebraic connectivity for a given $n$.}%
\label{fig:trees}%
\end{figure}

We used the software Nauty \cite{mckay1981practical} to generate all
non-isomorphic trees of maximal degree $d=3$ up to $n=23$ vertices (according
to Nauty, there are 565734 such trees with $n=23$). We then computed the tree
which maximizes $\lambda_{2}.$ The result is shown in Figure \ref{fig:trees}.
In all cases, the optimum tree was \textquotedblleft
well-balanced\textquotedblright\ in the sense that there was a central vertex
whose removal subdivides the tree into three nearly equal subtrees. The
maximizing tree was also unique. In the cases when $n=\frac{d\left(
d-1\right)  ^{K}-2}{d-2}$ (see Figure \ref{fig:trees}, $n=10$ or $n=22$), the
optimal tree appears to be the well-balanced Bethe tree whose algebraic
connectivity is well-studied \cite{molitierno2000tight, rojo2002spectrum,
rojo2006tight}, and is asymptotic to $\lambda_{2}\sim\frac{d}{d-1}\left(
d-2\right)  \frac{1}{n}$. These computations suggest that the bound
$\lambda_{2}\left(  T\right)  \leq2\left(  d-2\right)  \frac{1}{n}+O\left(
\frac{\ln n}{n^{2}}\right)  $ of Theorem \ref{thm:tree} is not optimal. We
propose the following optimal bound:

\begin{conj}
\label{conj:tree}Let $T$ be a tree with $n$ vertices and maximum degree $d.$
Then $\lambda_{2}\left(  T\right)  \leq\frac{d\left(  d-2\right)  }{d-1}%
\frac{1}{n}+O\left(  \frac{\ln n}{n^{2}}\right)  $ as $n\rightarrow\infty$ for
fixed $d.$
\end{conj}

In particular this conjecture is true for the well-balanced Bethe trees
mentioned above. A stronger version of this conjecture is

\begin{conj}
\label{conj:tree2}Let $T$ be a tree with $n=\frac{d\left(  d-1\right)  ^{K}%
-2}{d-2}$ vertices and maximum degree $d.$ Then its algebraic connectivity is
less than the algebraic connectivity of the well-balanced Bethe tree with $n$
vertices whose non-leaf vertices have degree $d$.
\end{conj}

We verified this conjecture using Nauty with $d=3$ and $n=10$ and $22$.

The bottleneck for improving Theorem \ref{thm:tree} into Conjecture
\ref{conj:tree} is Lemma \ref{Lemma:middle}. It states, roughly, that there is
a \textquotedblleft central vortex\textquotedblright\ whose removal subdivides
the tree into $d$ trees such that at least two have $\sim n/(2\left(
d-1\right)  )$ vortices. Indeed Conjecture \ref{conj:tree} is true for trees
that are \textquotedblleft well balanced\textquotedblright\ in the following sense:

\begin{prop}
\label{prop:tree}Suppose that a tree $T$ of order $n$ and maximal degree $d$
has a vertex whose removal subdivides $T$ into subtrees such that at least two
of the subtrees have at least $\left(  n-1\right)  /d$ vertices. Then
Conjecture \ref{conj:tree} is true.
\end{prop}

The proof of this proposition is identical to Theorem \ref{thm:tree}, except
that the bound $\frac{n-2}{2(d-1)}$ in Lemma \ref{Lemma:middle} gets replaced
by $\frac{n-1}{d}$, and therefore the prefactor $2\left(  d-2\right)  $ in
Theorem \ref{thm:tree} gets replaced by $\frac{d\left(  d-2\right)  }{d-1}.$

Proposition \ref{prop:tree} is applicable to all \textquotedblleft
maximal\textquotedblright\ trees in Figure \ref{fig:trees} as they happen to
be \textquotedblleft well-balanced\textquotedblright. But most trees are not
so well balanced. For example consider the following tree of 22 vertices:%

\[
\includegraphics[width=0.3\textwidth]{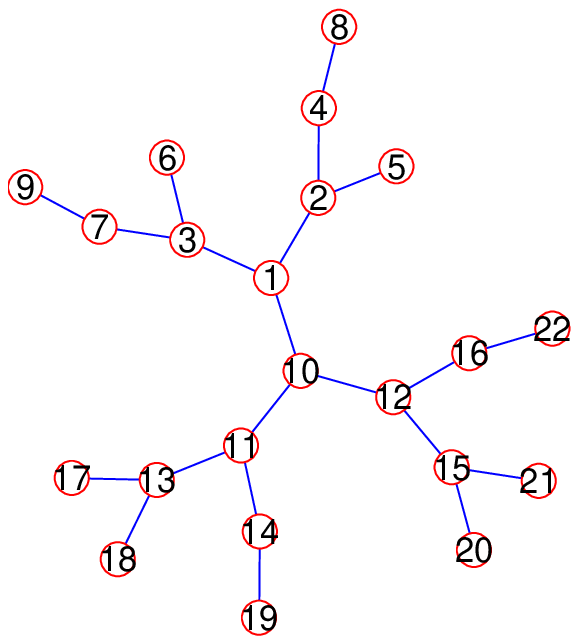}
\]
Proposition \ref{prop:tree} is not applicable to this tree:\ for example
removing vertex 10 results in three subtrees of size 9, 6 and 6 whereas
$(n-1)/d=7>6.$ Removing other vertices is even worse. Nonetheless for this
tree, $\lambda_{2}=0.0835$ which is smaller than $\lambda_{2}=0.0936$ of the
well-balanced tree of 22 vertices.

\begin{figure}[ptb]
\begin{center}
\includegraphics[width=1\textwidth]{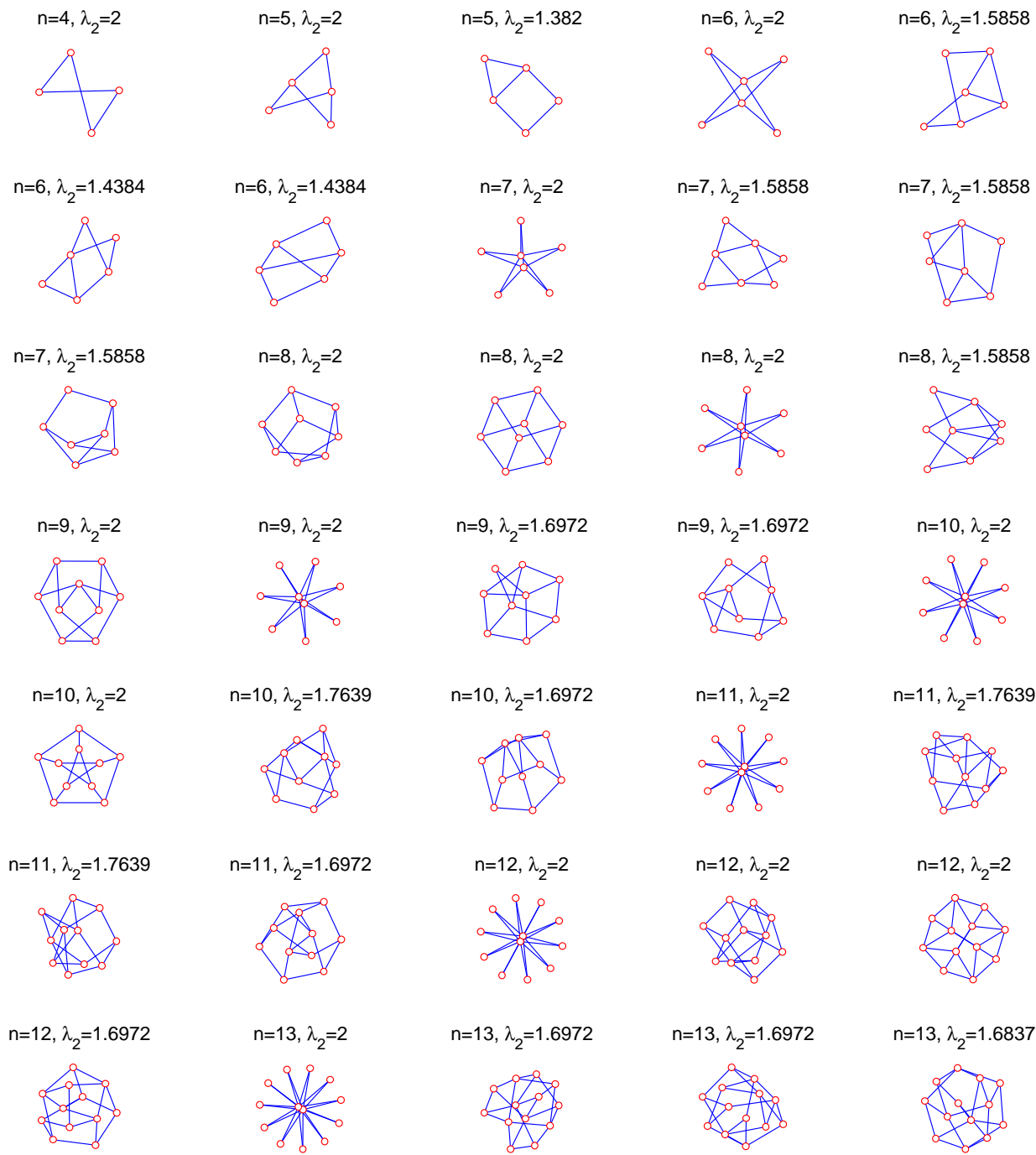}
\end{center}
\caption{Graphs with maximal algebraic connectivity with $m=2(n-2)$ edges.
Exhaustive search through all such graphs was done using Nauty program up to
$n=12$. For $n=13$, it was confirmed $\lambda_{2}\leq1.6972$ for graphs with
minimum degree 3 (graphs with minimum degree 2 have $\lambda_{2}$ at most
2n/(n-1)).}%
\label{fig:2n-4}%
\end{figure}

Consider again Conjecture \ref{conj:K2}, which states that $K_{2,n-2}$ has
optimal algebraic connectivity $\lambda_{2}=2$ among all graphs with
$m=2\left(  n-2\right)  $ edges. We used Nauty to exhaustively search through
all graphs with $m=2\left(  n-2\right)  $ edges and with $n$ up to 13, and
chose those with highest algebraic connectivity. The \textquotedblleft
winners\textquotedblright\ of this race are shown in Figure \ref{fig:2n-4}.
For all $n$ we tested, the highest connectivity $\lambda_{2}=2$ was attained
by the complete bipartite graph $K_{2,n-2},$ although depending on $n,$
several other graphs also had this connectivity. For example when $n=10,$
there are two graphs with $\lambda_{2}=2$: one is the Petersen graph and the
other is the complete bipartite graph $K_{2,8}.$ The number of graphs with
$m=2\left(  n-2\right)  $ edges seems to increase very fast with $n$:\ for
example Nauty returned $\approx2.7\times10^{7}$ non-isomorphic connected
graphs with $12$ vertices and $20$ edges whose minimum degree is 2, making it
prohibitively expensive to do an exhaustive search for bigger values of $n$
(we restricted the minimum degree to 2 because $\lambda_{2}$ is bounded by
$\frac{n}{n-1}d$ where $d$ is the minimum degree, and since we are only
interested in $\lambda_{2}$ well above 1)$.$ For $n=13$ (and $m=22$) we only
searched through graphs whose minimum degree is 3, of which there are were
about $1.6\times10^{6}$.\begin{figure}[ptb]
\begin{center}
\includegraphics[height=0.85\textheight]{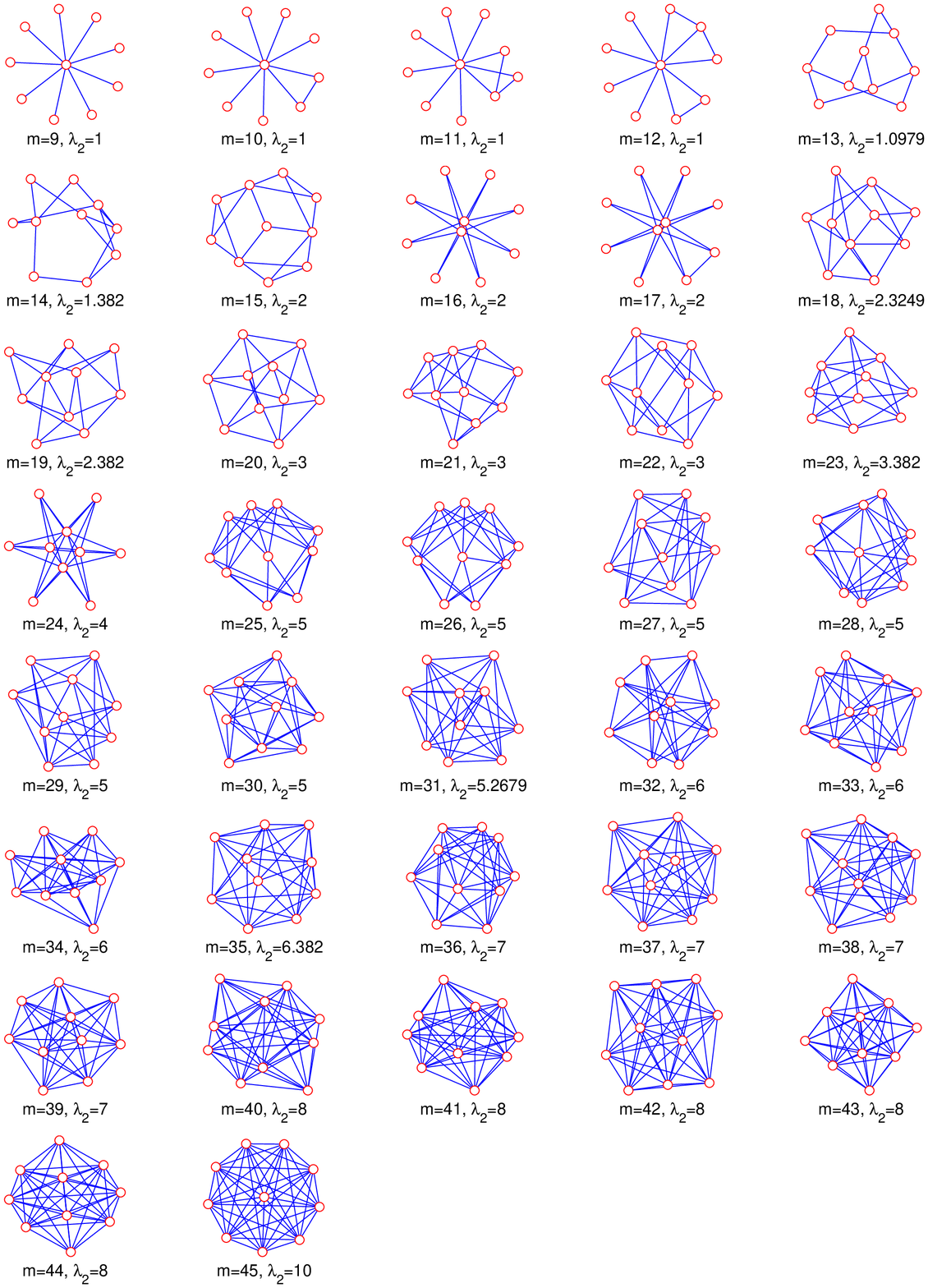}
\end{center}
\caption{ Some of the \textquotedblleft winning\textquotedblright\ graphs that
have maximum possible algebraic connectivity with $n=10$ vertices and $m$
edges, with $m$ as indicated. For some $m$, multiple maximizers exist but only
one is shown. }%
\label{fig:n10}%
\end{figure}

For larger $n,$ there appears to be a large jump between the maximum value
$\lambda_{2}=2$ and the next biggest value. For example with $n=13,$ the next
maximal value is 1.6972, with nearly uniform degree distribution (all vertices
have degree 3 or 4). The jump to the next $\lambda_{2}$ is much smaller
(1.6837). As far as we can tell, with the exception of $K_{2,n-2},$ all other
optimal or nearly-optimal graphs have vertices of degree either 3 or 4.

In Figure \ref{fig:n10} we list maximal graphs with $n=10$ and with varying
$m.$ Complete bipartite graphs with $m=b\left(  n-b\right)  $ are maximizers
with $\lambda_{2}=b,$ when $b=2,3,4,5$ and $n=10.$ It may be tempting to
generalize Conjecture \ref{conj:K2} as follows:\ 

\emph{Is it true that among graphs of }$n$\emph{ vertices and }$m=b\left(
n-b\right)  $\emph{ edges, the graph with the highest algebraic connectivity
of }$\lambda_{2}=b$\emph{ is attained by the complete bipartite graph
}$K_{b,n-b}$\emph{ when} $b<n/2$\emph{?}\ 

In fact, the answer is \emph{false}:\ it is known that for a random
$d-$regular graph, the expected algebraic connectivity is $\lambda_{2}\sim
d-2\sqrt{d-1}$ as $n\rightarrow\infty$ \cite{mckay1981expected,
broder1987second, friedman1991second}. Such graph has $m=dn/2$ edges. For
large $m,$ this corresponds to $b\sim d/2.$ Setting $d/2\sim$ $d-2\sqrt{d-1},$
we obtain that at least for $d\geq15$ and large $n,$ a random $d$-regular
graph has higher connectivity than $K_{b,n-b}$ with $b=d/2,$ with very high
probability. In other words, if $b\geq8,$ $K_{b,n-b}$ is not the maximizer of
$\lambda_{2}$ among the graphs of $b\left(  n-b\right)  $ vertices. This leads
to the following question:\begin{figure}[ptb]
\begin{center}
\includegraphics[width=0.45\textwidth]{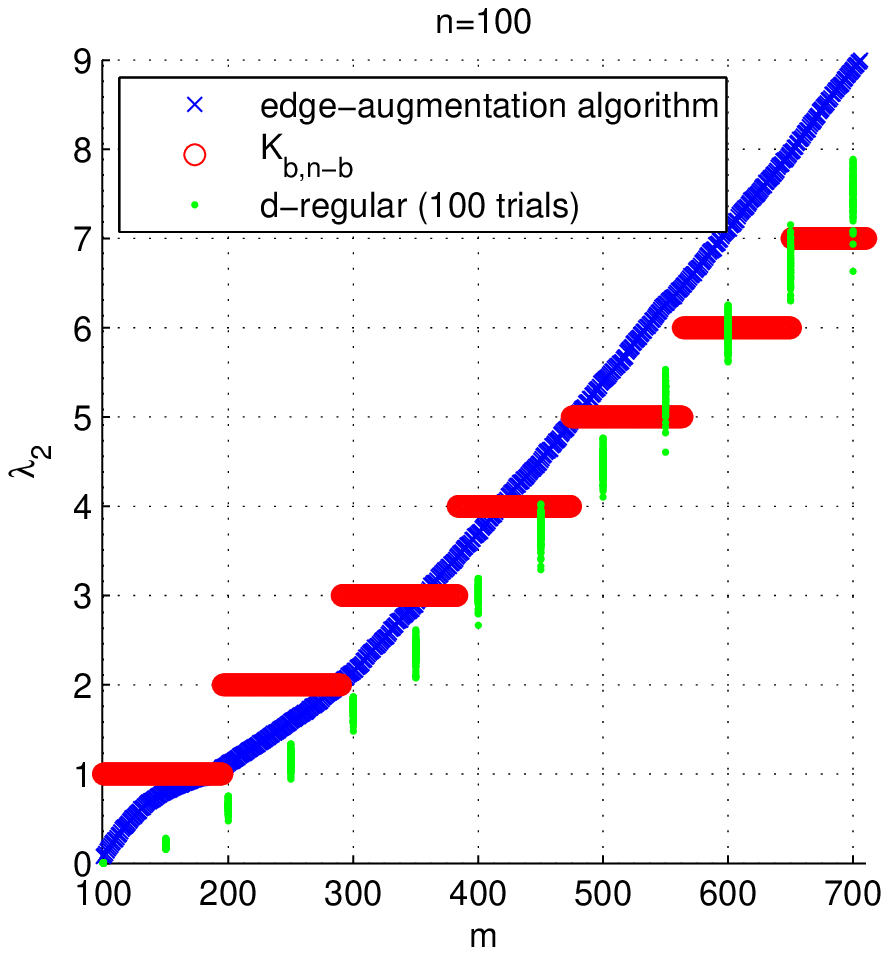}
\includegraphics[width=0.45\textwidth]{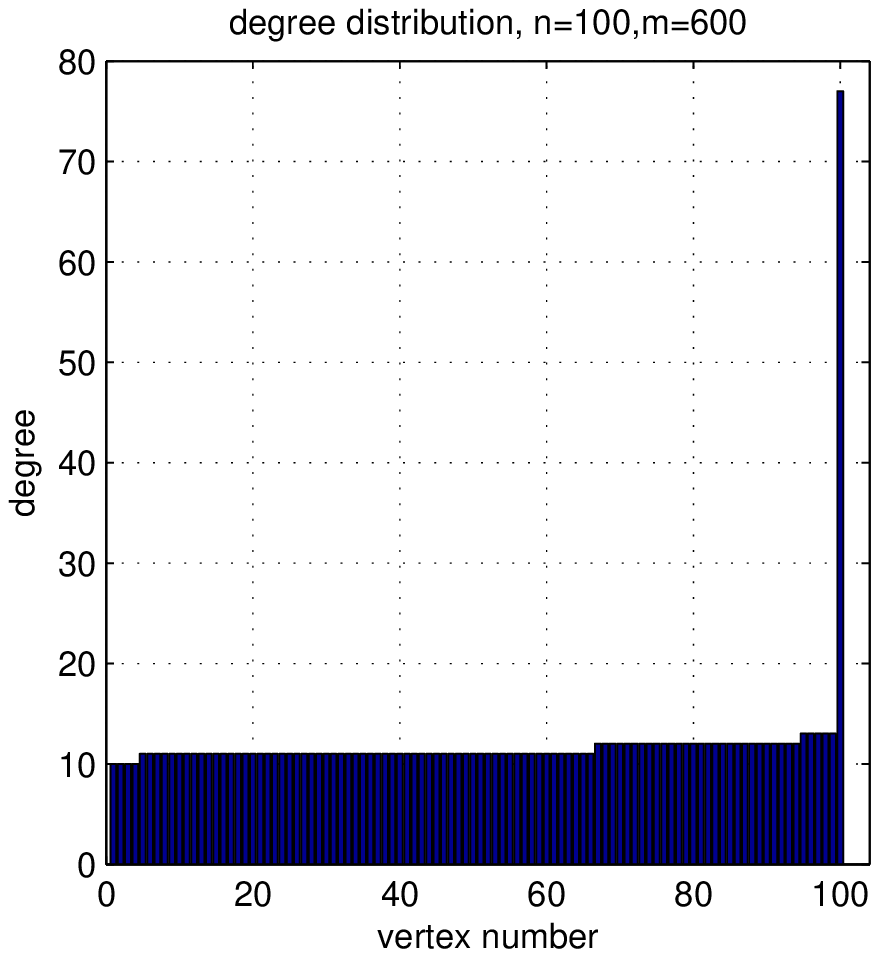}
\end{center}
\caption{Left: Comparison of algebraic connectivity obtained from
edge-augmentation algorithm versus $d$-regular graphs and bipartite complete
graphs $K_{b,n-b}$. For a given number of edges $m$, $b$ is taken to be the
largest integer such that $b(n-b)\leq m$ whereas $d=2m/n$. Right: The
distribution of degrees as obtained by the edge-augmentation algorithm with
$n=100,m=600$. In this regime the edge-augmentation beats the both the
$d$-regular graph (with $d=12$) and $K_{b,n-b}$ (with $b=6$). Note the
presence of a high-degree vertex. }%
\label{fig:compare}%
\end{figure}

\begin{openq}
Among graphs with $m=bn$ edges, what is the degree distribution for that
maximizes the algebraic connectivity, when $b\geq8,$ $b=O(1),$ and
$n\rightarrow\infty?$
\end{openq}

We speculate that this question could have implications for airline network
design. Most major US\ airlines utilize \textquotedblleft
hub-network\textquotedblright\ with several large airports serving multiple
smaller airports. This is similar to the complete bipartite graph $K_{b,n-b}.$
However the above results suggest that for airlines with more than 8 hubs, it
is may be better to switch to more uniform topology, with each airport having
roughly similar number of connections to others. Of course, there are many
other factors to consider for airlines, such as city size and popular travel
destinations, as well as the physical distance between cities. To what extent
the algebraic connectivity plays any role in airport design is unclear.

For values of $n>12$, exhaustive search is impractical and heuristic
algorithms to maximize connectivity need to be used. In
\cite{ghosh2006growing},\cite{wang2008algebraic}, the following
\textquotedblleft edge-augmentation\textquotedblright\ heuristic algorithm was
suggested to find graphs with $n$ vertices and $m$ edges having relatively
high algebraic connectivity:

\begin{enumerate2}
\item Start with an empty graph of $n$ vertices.

\item Compute the eigenvector $v$ corresponding to $\lambda_{2}(G).$

\item Find vertices $i,j$ for which $\left\vert v_{i}-v_{j}\right\vert $ is is
maximum. Add an edge $\left(  i,j\right)  $ to $G.$

\item Repeat steps 2 and 3 until the graph has $m$ edges.
\end{enumerate2}

The edge-augmentation gives better results when compared with $d$-regular
graphs, for the same number of edges $m=dn/2$. However for $b<5$ and with
$m=b(n-b)$, the complete bipartite graph $K_{b,n-b}$ has $\lambda_{2}=b$,
which is better than the edge-augmentation. On the other hand,
edge-augmentation overtakes both complete bipartite graph when $b>6,$ as well
as the $d-$regular graph with $d=2b>12$. This is illustrated in Figure
\ref{fig:compare} with $n=100$.

As mentioned in Remark \ref{remark:cubic}, the bounds of Theorem
\ref{thm:cubic} as well as in Lemma \ref{Lemma:Tk} are tighter than Nilli's
bound of $3-2^{3/2}\cos\left(  2\pi/D\right)  $. Our numerical investigations
suggest that this is true in general. We pose this as a conjecture.

\begin{conj}
\textbf{\label{conj:cubic}}Any cubic graph of diameter $D$ has algebraic
connectivity at most $3-2^{3/2}\cos\left(  \pi/D\right)  .$ Any cubic graph of
order $n=2^{K+1}-2$ has algebraic connectivity at most $3-2^{3/2}\cos\left(
\pi/K\right)  .$
\end{conj}

An $g-$cage is a cubic graph of girth $g$ with smallest possible number of
vertices. Motivated by the search for cages, many sophisticated techniques
have been developed for exhaustive enumeration of cubic graphs, especially for
those of high girth \cite{mckay1998fast, biggs1998constructions,
exoo2011computational, exoo2008dynamic}. For smaller $n,$ tables of cubic
graphs are available on the website House of Graphs,
http://hog.grinvin.org/Cubic. Upon checking these tables \emph{in every case
we checked, the maximizer for the algebraic connectivity of cubic graphs with
given number of vertices is also the graph that has the highest possible
girth. }Using the table we verified Conjecture \ref{conj:cubic} for $K$ up to
$3$ (when $n=14$). In the case $K=2,3,4$ and $6$ the conjectured bound is
actually attained by cubic graphs that have maximal possible girth as listed
in the following table.%

\begin{tabular}
[c]{l|l|l|p{4in}}%
$K$ & $n$ & $%
\begin{array}
[c]{c}%
\text{upper bound}\\
3-2^{3/2}\cos\frac{\pi}{K}%
\end{array}
$ & notes\\\hline
2 & 6 & 3 & Unique graph attains this bound. It has girth 4.\\\hline
3 & 14 & $1.58578$ & Nauty was used to verify that this bound is attained by a
unique cubic graph of girth 6, the Heawood Graph\\\hline
4 & 30 & $1$ & {The unique cubic graph of girth 8, the Tutte 8-cage. attains
this bound. All 545 cubic graphs with 30 vertices and with girth }$=7$ { have
algebraic connectivity strictly less than this.}\\\hline
5 & 62 & $0.71175$ & Of 27169 cubic graphs that have girth 9, none attain this
bound. Among them, maximum is $\lambda_{2}=0.603671$.\\\hline
6 & 126 & $0.55051$ & Tutte 12-Cage (girth 12) attains this.\\\hline
7 & 254 & $0.451675$ & ????
\end{tabular}
\newline

The maximal graphs listed above corresponding to $K=2,3,4,6$ all contain
$T_{K}$ as a subgraph; the case $n=14$ is shown in Figure \ref{fig:canonical}%
(d). However the maximizer graph for $K=5$ of girth 9 \emph{does not} contain
$T_{K}$. For $K\geq4,$ it is not known whether there are graphs with even
higher algebraic connectivity.

A complete list of graphs with 62 vertices and of maximal possible girth 9 was
kindly supplied by Brendon Mckay \cite{mckay-personal}. He computed it using
the program\ described in \cite{exoo2011computational}. The computation took
about 1000 machine hours and resulted in 27169 graphs of girth 9. Among these,
the maximal algebraic connectivity of $\lambda_{2}=0.603671$ was attained by a
single graph.

\section{Acknowledgements}

The author is grateful to Brendan McKay who generously supplied the complete
list of 27169 cubic graphs of girth 9 of order 62. I\ would like to thank
Braxton Osting for fruitful conversations. I\ also thank an anonymous referee
for suggesting a much better proof of Lemma \ref{Lemma:middle} than the
original revision of the paper, and numerous other suggestions. The author's
research is funded by NSERC discovery grant and NSERC\ accelerator grant.

\bibliographystyle{elsart}
\bibliography{graph}

\end{document}